# The Hard Truth about Soft Skills in Game Development

Benjamin Kenwright

School of Mathematics and Computing, Heriot-Watt University, Edinburgh, UK

## Abstract

This article explores the value and measurable effects of hard and soft skills in academia when teaching and developing abilities for the game industry. As we discuss, each individuals engagement with the subject directly impacts their performance; which is influenced by their 'soft' skill level. Students that succeed in mastering soft skills earlier on typically have a greater understanding and satisfaction of the subject (able to see the underlying heterogeneous nature of the material). As soft and hard skill don't just help individuals achieve their goals (qualifications), they also change their **mindset**. While it is important to master both hard and soft skills, often when we talk about the quality of education (for game development); the measure is more towards quantitative measures and assessments (which don't always sit well with soft skills). As it is easy to forget, in this digital age, that 'people' are at the heart of video game development. Not just about 'code' and 'technologies'. There exists a complex relationship between hard and soft skills and their dual importance is crucial if graduates are to succeed in the game industry.

## 1 Introduction

Soft and hard skills have long been an area of debate in technical subjects over the years [Choudary 2014; Kenwright 2016a]; especially when it comes to employability [Succi and Canovi 2020]. Of course, most everyone would agree, to some extent, you need both to succeed in the video game industry. This article aims to inspire those who teach game development subjects - to appreciate and understand the diversity and complexity of the skills necessary for students to succeed in the industry.

The video game industry (technical focused areas), is one of the most competitive and fast-growing sectors. Even though the demand for **skilled developers, designers, project managers and marketers** is still growing, the industry only picks the most skilled. And when we talk about skills, what kind of skills are we referring to? Which skills are in most demand (and how are they measured)? [Kenwright 2016a] As online tools and methods of learning are becoming more prominent; are certain soft skills being lost? [Kenwright 2020]. For instance, social factor of casual debate and discussion in an unstructured open environment? Then again, do flexible modes of learning provide new models of learning that are able to take leaning above and beyond? (e.g., VR/XR [Kenwright 2019] which could let students experience complex stressful situations to help them develop and grow their abilities to cope/manage).

**Soft skills in the video game industry is a necessity, not an option.**

## 2 Defining skills

For completeness, we define the details behind what hard and soft skills mean here. Of course, be aware that hard and soft skills are very rarely mastered exclusively, for instance, hard skills are often learned in a sphere that develops other skills passively (e.g., communication and problem-solving).

**Hard skills** are technical knowledge or training that individuals gain through taught material (education), life experiences or career. They are practical and often relate to mechanical, information technology, mathematical or scientific tasks. Some examples include knowledge of programming languages, design programs, hardware equipment or tools.

**Soft skills** are personality traits and behaviors. Unlike hard (technical) skills soft skills are not about the knowledge individuals possess but rather the **behaviors**, they display in different situations. They are a complex combination of people skills, social skills, communication skills, character or personality traits, attitudes, mindsets, career attributes, social intelligence and emotional intelligence.

Following on from this, we discuss popular methods for developing and assessing these skills (e.g., programming assignments to team projects).

## 3 Developing skills

When we develop hard skills, it is through education (books and knowledge) or practice-based tasks (typing in exercises/code), while developing soft skills is through more 'personal' experiences.

Improving hard skills or acquiring new ones is uncomplicated way to add measurable value to an individuals resume. Developing these **hard skills takes time, practice and applied application**. It is difficult to become proficient in a hard skill until the individual has actually used it for a project or solved a problem/task with it. Straightforward at university to roll-out hard skills in small assessments or through personal projects where the stakes are not as high. Once individuals have thoroughly practiced their newly acquired skills, they can are able to demonstrate and show them off with confidence.

Assessing an individuals soft skills, however, takes some introspection. For instance, an individual may think of themselves as a "collaborative leader", until colleagues and peers provide feedback (assessors and individuals themselves have a limited view of a person's soft skills). There are also technological solutions for assessing soft skills today. For instance, 'personality tests' which give the individual an overview of their strongest and weakest interpersonal traits. As well as the hard skills, **soft skills can be thought and trained**. Suitable assessments (e.g., peer review), reflective material, mentoring sessions and even talking with peers helps individuals learn and improve their soft skills.

## 4 Measuring skills

Hard skills are measurable and can be described using numerical or yes/no criteria. On the other hand, **soft skills are often intangible or hard to quantify** and are usually described with qualitative scales. It can be hard to prove proficiency in soft skills, but they are necessary (as they create a positive and functional foundation for the individual).

Developing assessments and measurements for soft skills is more **complicated** than hard skills, **but not impossible**. For instance, group projects, peer assessments, presentations, discussions and code reviews with feedback provide soft-skill insights that would be missed through traditional assessment methods (individual reports or exams) [Kenwright 2016b]. Importantly, these alternative ways of assessing students can challenge students on a range of levels through activities that building their understanding and mindset.

**Qualitative vs Qualitative** Just to emphasise, soft skills are often intangible or hard to quantify and are usually described with **qualitative scales**. Yet, hard skills are easy to identify and evaluate, not just in academia, but also by industry, through resumes, portfolios, job-related assignments and role-specific interview questions.

When demonstrating or wanting to assess an individuals soft skills - these are done by asking situational and behavioral questions; soft skill questions and tests that take into account an individual's overall **personality characteristics**. Sometimes, looking at the individuals interests, participation in societies, open source projects, events, fund-raisers, hackathons

Soft skills are 'personal' habit traits that shape how individuals work (on their own and with others). Brief list gives an idea of the diversity and the challenges in developing these areas:

- Integrity
- Dependability
- Effective communication
- Open-mindedness
- Teamwork
- Creativity
- Problem-solving
- Critical-thinking
- Adaptability
- Organization
- Willingness to learn
- Empathy

Hard skills are technical knowledge or training that individuals gain through life experiences (including education and career development). The few hard-skill examples are easy to quantify and measure (easy to see how they would be assessed and taught on a course):

- Programming languages (Python, C, Java, ..)
- User interface and interaction 'design' methods
- Database (data) management
- Software packages/tools/suites
- Network security
- Game marketing/sales
- Statistical analysis
- Data mining
- Mobile development
- Team/project management
- Storage system and management
- Hardware architecture

gives insights into the 'type' of person (i.e., soft skills); which are not 'taught' to students.

## 5 Skills for the future game industry

Soft and hard skills are both important strengths and the key not only for building effective video game developers but for thriving any constantly evolving workplace. Soft skills and hard skills go hand in hand. There is no question or hesitation. There should be no debate on which is more important. For instance, imagine an employer looking for a game developer, hard skills are necessary for this role to include knowledge of specific programming languages (e.g. shaders or C++), frameworks and tools. On the other hand, useful soft skills for everyday work are communication, collaboration, problem-solving attitude and time management abilities to be a successful game developer. There will always be people who build groundbreaking products and produce unparalleled content, but there will also be people who will need to manage and lead these endeavors. A video game studio (or any organisation) cannot be successful with only hard-skilled or only soft-skilled employees. Finding harmony between the two categories of skills and how they complement each other is crucial. As graduates leave academia and enter the workforce; it's never been more important to for them to be aware of the most in-demand skills for the future. While the video game industry continues grow and expand; the educational sector has an opportunity to help reassure their students that they are enhancing their capabilities not just for 'now' and based on 'hard' facts but the ability to meet tomorrows needs (flexible skillset).

Technology is at the forefront of video games (and every industry), so academia needs to nurture this creative and innovative thread in students if they're to survive. You could say, creativity is the **most in-demand soft skill of the future**. Increasing requirement for complex problem-solving skills and the ability for individuals to think critically [Berger and Frey 2015]. In the constantly changing environment, it's crucial that individual's are agile and able to pivot easily, have a **growth mindset**. The ability to continually improve so they can be great at anything when they put their mind to it. Having a growth mindset involves continually being on the lookout for the latest trends and innovations that can help them be, and perform, better.

In video game organisations, digitally-capable individuals should be able to outperform non-digital individuals (as they're more well informed). Hence, organisations, needs to take a broader view of what it means to have **leadership potential** (i.e., not just about people/leadership skills if they're to be successful in their role).

## 6 Ability to Communicate

Communication is such an important skill. Generally people often under-appreciate the problems caused by miscommunication (both written and oral) and is a core skill that sits at the top of every industry and area of learning. Any miscommunication leads to complication. Clearly, the problems of poor communication go far beyond academic, social and industry barriers. **Poor communicators** tend to believe that quantity is more important than quality. But skilled communicators pay careful attention to all points, and closely find and understand the details at the heart of information. When they do write or talk, they adjust their tone and style to the audience. Their colleagues feel understood and respected, no matter the situation. Without the ability to communicate skillfully, other important skills, like programming, problem-solving and team-work become impossible. Now more than ever, academia should be encouraging and developing good communication skills from the start. The most common method, is that students work in teams and groups to develop solutions while investing time and energy into cultivating the necessary communication skills to succeed. From interpersonal communication and teamwork to adaptability, creativity, and attitude, **communication skills are not industry-specific and are not always emphasized in professional development as much as technical capabilities**. Video game industry is changing, no longer judge a candidate or graduate based on their 'qualifications' - they recognize that an individuals skills stretch beyond the classroom (and are essential for success).

## 7 Student Demand

Since hard skills are more specific and measurable; and are often key requirements for game development jobs (e.g., technical skill); this means students and academic courses tend to lean more towards these (courses that teach and assess hard-skills vs soft-skills) [Kenwright 2016a]. Since soft-skills are highly dependent on an individuals personality and behavioural qualities - these are not able to be 'changed' or modifies easily through 'taught' methods (e.g., working in a group doesn't make a student good at group work). However, these insights and experiences, help students understand their limitations and areas for improvement. For subjects, such as game development, it is possible to take advantage of interactive and social elements (gamification) to encourage and nurture other skillset [Kumar and Lightner 2007]. Courses need to educate and prepare students on the wider context (i.e., more than just learning technical concepts) - must also 'teach' students soft-skills though carefully developed curricula. Benefits for both students and the universities (not whimper to what students want but what they need).

## 8 Conclusion/Discussion

Soft skills allow individuals to connect with others, convey ideas, and learn from those around them. Are universities adequately developing and training students with soft skills that will land them their dream job in the game industry? Soft skills are undersold. Even though it is vital to invest time and energy in teaching the technical skills students need, **soft skills are just as critical**. The majority of game development (if not all) jobs involve teamwork; and soft skills are more important than ever.

Disseminating the relationships between hard and soft skills within courses; helps identify metrics and meanings behind other measures, such as student satisfaction and engagement. While this article has focused on game development subjects, the results and assumptions presented here are important and transferable to other areas.